\documentstyle[osa,aps,prl,epsfig]{revtex}
\begin{document}
\twocolumn[\hsize\textwidth\columnwidth\hsize\csname
@twocolumnfalse\endcsname
\title{
Gate-Induced Mott Transition}
\author{Hyun-Tak Kim$^{\ast}$, B. G. Chae, D. H. Youn, S. L. Maeng, K. Y. Kang}
\address{Telecom. Basic Research Lab., ETRI, Daejeon 305-350, Korea}
\date{May 27, 2003}
\maketitle{}
\begin{abstract}
For a strongly correlated material, VO$_2$, near a critical
on-site Coulomb energy $U/U_c$=1, the abrupt Mott metal-insulator
transition (MIT) rather than the continuous Hubbard MIT is
observed by inducing internal optical phonon-coupled holes (hole
inducing of 0.018$\%$) into conduction band, with a gate field of
fabricated transistors. Observed gate effects, change of the MIT
drain-source voltage caused by a gate field, are the effect of
measurement due to inhomogeneity of channel material and is an
average over the measurement region of the true gate effect based
on the large conductivity (or effective mass) near the MIT
predicted by the Brinkman-Rice picture. A discontinuous gate
effect such as digital is observed, which is a characteristic of
the transistor and a possible condition of a very high-speed
switching transistor.
\\
\end{abstract}
]
In a strongly correlated system, a metal-insulator transition
(MIT) near a critical on-site Coulomb energy, $U/U_c$=1, has long
been controversial in terms of whether the transition is Mott's
abrupt (or first-order) MIT or Hubbard's continuous (or
second-order) MIT $[1-4]$. This is because it is unclear whether
experimental observations of the abrupt MIT (Mott transition)
follow Mott's prediction, although a first-order MIT with
temperature was observed by McWhan $et~al.~[5]$. Rather than the
abrupt MIT, Boriskov $et~al.~[6]$ and Kumai $et~al.~[7]$ measured
the nonconduction-conduction transition (NCT) with an electric
field for VO$_2$ and an organic material, respectively. Oka
$et~al.~[8]$ found the NCT with an electric field could be
described in terms of a universal Landau-Zener quantum tunnelling
through a theoretical consideration based on the Hubbard model.
Newn $et~al.~[9]$ regarded the NCT in a Mott-Hubbard insulator as
the Hubbard MIT for a transistor based on the NCT. However, the
Mott MIT, which differs from the NCT, has not been observed for a
very low doping of charges (hole content of 0.018$\%$ for VO$_2$),
as predicted by Mott $[1]$. The very low doping is a decisive key
determining an order (first or second) of two kinds of the MIT.

An abrupt MIT breaks down an energy gap between sub-bands in a
main band. The energy gap is formed by a strong correlated on-site
Coulomb energy. The abrupt MIT in a strongly correlated metal with
an electronic structure of one electron per atom was theoretically
demonstrated by Brinkman and Rice; this is called the
Brinkman-Rice (BR) picture $[10]$. The abrupt MIT with band
filling was also developed through extension of the BR picture by
Kim $[11,12]$. The extended BR picture was based on a fractional
charge justified by means of measurement in an inhomogeneous
metallic system $[12]$.

In this letter, we observe the abrupt MIT (or Mott transition),
inducing internal optical phonon-coupled hole charges (hole
content of 0.018$\%$) into conduction band with a gate field of a
fabricated field-effect transistor. This reveals a difference
between the abrupt MIT and the continuous MIT. Note that
artificial hole doping of 0.018$\%$ is not possible other than the
method applying the gate field.

In the extended BR picture $[11,12]$, the effective mass,
$m^{\ast}$, of a carrier is given by
\begin{eqnarray}
\frac{m^{\ast}}{m} {\equiv}\frac{1}{1-(U/U_c)^2} =
\frac{1}{1-{\kappa^2}{\rho^4}} ,
\end{eqnarray}
where $m$ is the bare electron mass, $U/U_c$ is
${\kappa\rho^2\ne}1$, $\kappa$ is the strength of Coulomb energy
between carriers when $\rho$=1, and 0$<{\rho\le}1$ is band
filling. When ${\rho\ne}1$, Eq. (1) is well fitted in a real
metallic system and the effect of measurement (or average) for an
inhomogeneous system $[12]$. An electric conductivity,
$\sigma~{\propto}~(m^{\ast}/m)^2$ $[1]$.

The material at $\kappa\rho^2$=1 in Eq. (1) can be assumed as a
paramagnetic insulator (or Mott insulator). The metal at a
critical $\rho$ value (=$\rho'$) of just below $\rho$=1 shows the
best metallic characteristics $[13]$. The MIT from a metal at both
$\rho'$ and $\kappa$=1 (${\kappa}{\rho'^2}{\ne}$1) to the
insulator at both $\rho$=1 and $\kappa$=1 (${\kappa}{\rho^2}$=1)
is abrupt (or a jump); this is an idea for observing the Mott
transition near $U/U_c$=1. Holes corresponding to the difference
(critical hole content, ${\triangle\rho'}=1-\rho'$) between
${\rho}'$ and ${\rho}$=1 are induced into a conduction band having
electrons by a gate electric field; this is regarded as the
decrease of the Coulomb energy $[12]$. Then, the energy gap breaks
down and the metallic system becomes inhomogeneous because of the
induced holes $[12]$. The number of the induced holes can be
regarded as $n_c~{\approx}~3{\times}10^{18}~cm^{-3}$, predicted by
Mott from $n_c^{1/3}a_0\approx$0.25 $[1]$. Here, $a_0$ is the Bohr
radius and $n_c$ corresponds to about 0.018$\%$ of the number of
carriers in the half-filled band, when one electron in the cell
volume, 59.22$\times$10$^{-24}$ cm$^3$, of VO$_2$ is assumed;
${\triangle\rho'}$=0.018$\%$. Further, a gate electric field of a
transistor induces holes in optical phonon-coupled-hole levels
$[14,15]$ in a Mott insulator, a channel material, into conduction
band $[16]$. The process in which optical phonon-coupled holes
change to carriers has been revealed $[15]$. The hole levels are
attributed to impurities such as oxygen deficiency, which
indicates that VO$_2$ is inhomogeneous, as proved experimentally
by Kumai $et~al.~[7]$.

We fabricate transistors to observe the Mott transition on the
basis of the above theory. The schematic diagram of the transistor
is shown in Fig. 1. Thin films of the Mott insulator, VO$_2$, with
a sub-energy gap of about 1~$eV$ in the $d$-main band $[17]$ have
been deposited on Al$_2$O$_3$ substrates by laser ablation. The
thickness of the VO$_2$ film is about 900$\AA$. The resistance of
the film decreases with increasing temperature and shows an abrupt
MIT at a transition temperature, $T_{tr}$=340 K (68$^{\circ}$C)
(Fig. 2a). This is the same as that measured by Borek
$et~al.~[18]$. The decrease of the resistance up to 340 K
indicates an increase of hole carriers, and two kinds of electron
and hole carriers coexist near $T_{tr}$=340 K (Fig. 2b). From 332
to 340 K, the number of carriers is not clear because of mixing of
electrons and holes. We speculate that the number of hole carriers
can be $n_c~{\approx}~3{\times}10^{18}~cm^{-3}$ at $T_{tr}$=340 K
on the general basis that an exponential decrease of the
resistance with temperature in semiconductor physics indicates an
exponential increase of carriers. In the metal regime above 340 K,
carriers are electrons (Fig. 2).

Gate insulators, amorphous Ba$_{0.5}$Sr$_{0.5}$TiO$_3$ (BSTO),
Si$_3$N$_4$ and SiO$_2$ were used. BSTO and Si$_3$N$_4$ were
deposited on the VO$_2$ film at a VO$_2$ surface temperature,
400$^{\circ}$C and 150$^{\circ}$C, respectively. The thickness and
the dielectric constant of the BSTO and Si$_3$N$_4$ films were
about 1200$\AA$, 43 and 2000$\AA$, 7, respectively. Transistors of
channel length, $L_{ch}$= 3${\mu}m$, and gate width,
$L_w=50{\mu}m$, were fabricated by lithography processes. The gate
width of a transistor based on Si substrate is $L_w=25{\mu}m$.
Au/Cr electrodes were prepared for Ohmic contact. Characteristics
of the transistors were measured by a precision semiconductor
parameter analyzer (HP4156B). To protect transistors from excess
current, the maximum current was limited.

Figure 3a shows the drain-source current, $I_{DS}$, vs the
drain-source voltage, $V_{DS}$, characteristics of transistor 1
with a gate insulator of BSTO. The measured gate current,
$I_{GS}$, between the gate and the source is an order of
10$^{-13}~A$ at gate voltages of $V_G$=0, -2 and -10V, which
indicates that there is sufficient insulation between the gate and
the source. Fig. 3b shows a NCT below the MIT-$V_{DS}$ of point A
of curve 1 measured by an applied electric field between the
source and the drain at V$_G$=0V. This was observed by using a
two-terminal structure by Boriscov $et~al.~[6]$ and Kumai
$et~al.~[7]$ who used an organic Mott-insulator. The two groups
suggested that the NCT occurs due to an applied field $[6]$ and an
induced current $[7]$, not an increase of sample temperature due
to leakage current. The abrupt MIT of curve 1 has been measured
more than 1,500 times without breakdown. $I_{DS}$ follows the
Ohmic behavior up to $V_{DS}\approx$12V, but shows nonlinear
electric conduction in the total regime below the MIT-$V_{DS}$ of
point A (Fig. 3b). The nonlinear conduction behavior is regarded
as semiconducting behavior due to the increase of hole carriers by
Zener's impact ionization, as observed by the Hall effect (Fig.
2b). It was revealed through a theoretical consideration that the
Ohmic behavior is described in terms of a universal Landau-Zener
quantum tunnelling [8]. The NCT is an insulator-semiconductor
transition. We suggest that the abrupt MIT at point A occurs when
the number of hole carriers produced by impact ionization becomes
the number of $n_c$ predicted by Mott. The semiconduction in Fig.
3a is regarded as the doping process in which $n_c$ (or
$\triangle\rho'$) of holes are induced by electric field.
Moreover, at the jumped point in curve 1, the current density is
about $J$=3$\times$10$^5$ A/cm$^2$, which is current collective
motion observed in metal.

Curves 2 and 3 are $I_{DS}$ vs $V_{DS}$ characteristics measured
at gate voltages, V$_G$= -2 and -10$V$, respectively. The abrupt
MITs also occur at transition points B and C. The sharp
transitions indicate that transistor 1 was well fabricated. The
abrupt MIT at point B, which is caused by the induced charges of
$\triangle\rho'$, occurs suddenly at $V_G$=-2V, such as digital,
as indicated in Eq. (1). This is the most unique characteristic of
this transistor. The gate effect, a change of the MIT-$V_{DS}$
caused by a gate field, is small, which is that the channel
material in the measurement region is inhomogeneous (see, $[12]$).
When only homogeneous region is measured, the true gate effect
becomes maximum due to the maximum conductivity (or the maximum
effective mass) near the MIT as indicated in the BR picture;
MIT-$V_{DS}\propto(J/\sigma)~\rightarrow$ 0. The gate effect
increases as homogeneity of the VO$_2$ film increases. Thus, the
observed gate effect is an average of the true gate effect over
the measurement region. $n_c$ (or $\triangle\rho'$) may be
attained by the gate effect and the impact ionization on the
ground of the large transition $V_{DS}$.

Figure 3c shows current-voltage characteristics of transistor 2
with a gate insulator of an amorphous Si$_3$N$_4$. The gate-source
current is $I_{GS}\approx$3.6${\times}$10$^{-12}$A through the
Si$_3$N$_4$ at $V_G$=-2V and $V_{DS}$=0.1V. An off-current is
$I_{DS}\approx$1.3${\times}$10$^{-7}$A at point D. The abrupt MITs
occur at point F of $V_{DS}$=13V (or $E$=4.3MV/m) at $V_G$=0V and
point G of $V_{DS}$=9V (or $E$=3MV/m) at $V_G$=-2V. The gate
effect at point G at $V_G$=-2V is similar to that of transistor 1;
there is no gate effect at -2V$<V_G<$0. The discontinuous gate
effect is a possible condition of a very high-speed transistor.
Namely, transconductance related to a switching speed can be
regarded as maximum.

Figure 3d shows $I_{DS}$ vs $V_{DS}$ near the abrupt MIT of
transistor 3 with a gate insulator of SiO$_2$. Its structure is a
VO$_2$/SiO$_2$/WSi/Si substrate. Its characteristics are given as
follows. First, the gate effect at $V_{DS}$=14V and
$V_{gate}$=-10V, is due to hole inducing of
$\triangle\rho'=0.018\%$; this is the abrupt MIT with band filling
near $U/U_c$=1 in Eq. (1) and due to the jump of the gate voltage.
Second, the MIT-$V_{DS}$ increases with an increasing negative
gate voltage (or field), which is the decrease of the conductivity
(or the effective mass in Eq. (1)); this is due to the continuous
change of the gate voltage. Moreover, the increase of the inducing
hole content greater than the critical content,
$\triangle\rho>\triangle\rho'$, decreases $\rho$ and the
electrical conductivity, $\sigma$, because of
$\sigma~{\propto}~(m^{\ast}/m)^2$; $\sigma$ is maximum at
$\triangle\rho'$. From current density of $J$=${\sigma}E$, an
electric field, $E$, increases with a decreasing $\sigma$ to
constant $I_{DS}({\propto}J_{DS}$), as shown at transition points
in Figs. 3a, 3c and 3d, $vice~versa$; $E{\propto}V_{DS}$. We also
observed that the MIT-$V_{DS}$ decreases with an increasing
positive gate voltage. Third, $I_{DS}$ in the metal regime over
$I_{DS}$=2mA shows Ohmic's law which differs from the behavior in
Fig. 3b. At the jumped point, current density is
$j$=0.9$\times$10$^5$ A/cm$^2$, which is current collective motion
measured in metal. Thus, Fig. 3d follows the behavior of Eq. (1)
when $\kappa$=1.

We suggest conditions of a good transistor fabrication by
comparing transistors 1 and 2. The off-current ($I_{DS}$ at a very
low $V_{DS}$ and $V_g$=0V) and the MIT-V$_{DS}$ values of
transistor 1 are lower and higher, respectively, than those of
transistor 2; $V_{DS}$=20.8V (or $E$=7MV/m) of point A and
$V_{DS}$=13V (or $E$=4.3 MV/m) of point F. The smaller V$_{DS}$
results from the higher off-current arising from an oxygen
deficiency of VO$_2$ when the gate insulator is deposited. The
off-current is caused by the excitation of the optical
phonon-coupled holes. When the number of total holes in the hole
levels is given by $n_{tot}=n_b + n_{free}$, where $n_b$ is the
number of optical phonon-coupled holes and $n_{free}$ is the
number of holes freed from the levels, $n_b$ decreases with
increased $n_{free}$, because $n_{tot}$ is constant. In oxide
materials, $n_{tot}$ is about 5.5${\times}$10$^{18}cm^{-3}$ which
corresponds to 0.034$\%$ to $d$-band charges $[14,15,19]$. The
larger off-current is attributed to the increase of $n_{free}$.
For the abrupt MIT, ${\triangle}n{\equiv}n_c-n_{free}$=0 should be
satisfied, where $n_c~{\approx}~3{\times}10^{18}~cm^{-3}$, as
predicted by Mott. Hence, the decrease of ${\triangle}n$
contributes to the reduction of the MIT-$V_{DS}$ (Fig. 3e). In
Fig. 3c, the smaller MIT-$V_{DS}$ of transistor 2 is due to the
smaller $n_{induced}$=${\triangle}n$ induced by the gate electric
voltage (field) than that of transistor 1. For a good transistor,
off-current can be decreased when the deposition temperature or
oxygen content of VO$_2$ film is slightly increased. In addition,
when gate length is less than 100nm, $V_{DS}$ is much less than
1V.

Figure 3f shows the magnified part below $V_{DS}$=1.5V of the
curves in Fig. 3a. A current signal such as noise in curve 1
without $V_G$ is measured when the resistance of the VO$_2$ film
between the source and the drain is large. When $V_G$ are applied,
curves 2 and 3 show a high gain current of about 250 times at
$V_{DS}$=0.3V, which is the gate effect observed in a
semiconductor transistor. The high gain represents a significant
difference between this transistor with the abrupt MIT and the
Mott transistor developed with a Mott-Hubbard insulator,
Y$_{1-x}$Pr$_x$Ba$_2$Cu$_3$O$_{7-\delta}$, by Newn's $et~al.~[9]$
Although they regard a transition for the Mott transistor as the
Hubbard MIT, the transition is the NCT. Their transistor is not
the Mott transistor.

In conclusion, the abrupt Mott MIT near $U/U_c$=1 is first
observed by inducing of internal holes of 0.018$\%$ with gate
fields, while the continuous Hubbard MIT does not exist in Mott
and Mott-Hubbard insulators. Generally, the reason that the
continuous MIT is observed in strongly correlated systems
including high-$T_c$ superconductors is because the continuous
behavior in both Eq. (1) and Fig. 3d is observed when a doped
content, $\triangle\rho>\triangle\rho'$. The measured gate effects
predicted by Eq. (1) due to inhomogeneity of channel material is
an average of the true gate effect. Furthermore, the transistor
developed here is a true Mott transistor without short channel
effects proposed in metal-oxide-semiconductor field-effect
transistors and will be very useful for nano-devices.

We thank Dr. Soo-Hyeon Park at KBSI for Hall-effect measurement,
Dr. Gyungock Kim for valuable discussions on the Zener transition,
and Dr. J. H. Park for fabrication of Si$_3$N$_4$ films by using
CVD. HT Kim, the leader of this project, developed the concept,
and wrote the paper. BG Chae and DH Youn deposited VO$_2$ and BSTO
films, performed the transistor fabrication process, and measured
$I-V$ characteristics. KY Kang prepared the laser-ablation and
lithography equipment and generated this project with HT Kim. SL
Maeng evaluated transistor characteristics, the shielding
measurement system, and Si$_3$N$_4$ film fabrication.

\begin{figure}
\vspace{-1.1cm}
\centerline{\epsfysize=6.0cm\epsfxsize=8cm\epsfbox{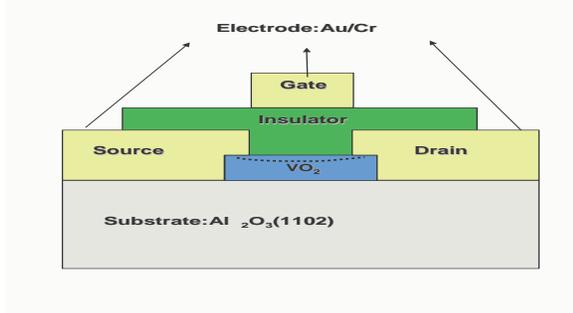}}
\vspace{-1.3cm} \caption{Schematic diagram of a transistor. A dot
line between VO$_2$ and insulator is channel. The gate insulators
are an amorphous Ba$_{0.5}$Sr$_{0.5}$TiO$_3$ for transistor 1 and
an amorphous Si$_3$N$_4$ for transistor 2.}
\end{figure}


\begin{figure}
\vspace{-0.7cm}
\centerline{\epsfysize=8cm\epsfxsize=8cm\epsfbox{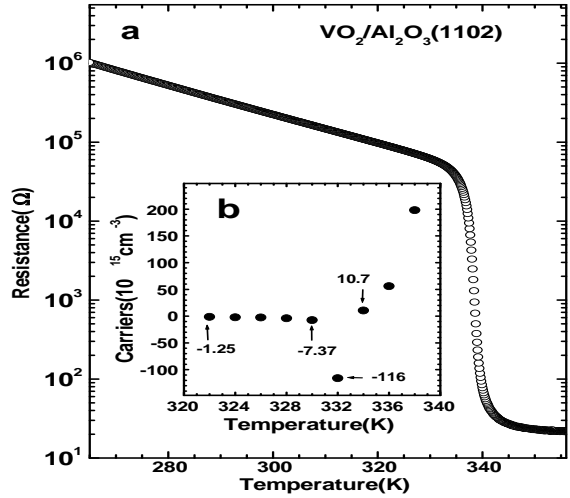}}
\vspace{0.3cm} \caption{${\bf a}$, Temperature dependence of the
resistance of a VO$_2$ film. ${\bf b}$, The number of carriers
measured by Hall effect. A change of carriers from hole to
electron is shown at 332 K. The minus sign indicates that carriers
are holes.}
\end{figure}


\begin{figure}
\vspace{-0.5cm}
\centerline{\epsfysize=14cm\epsfxsize=8.4cm\epsfbox{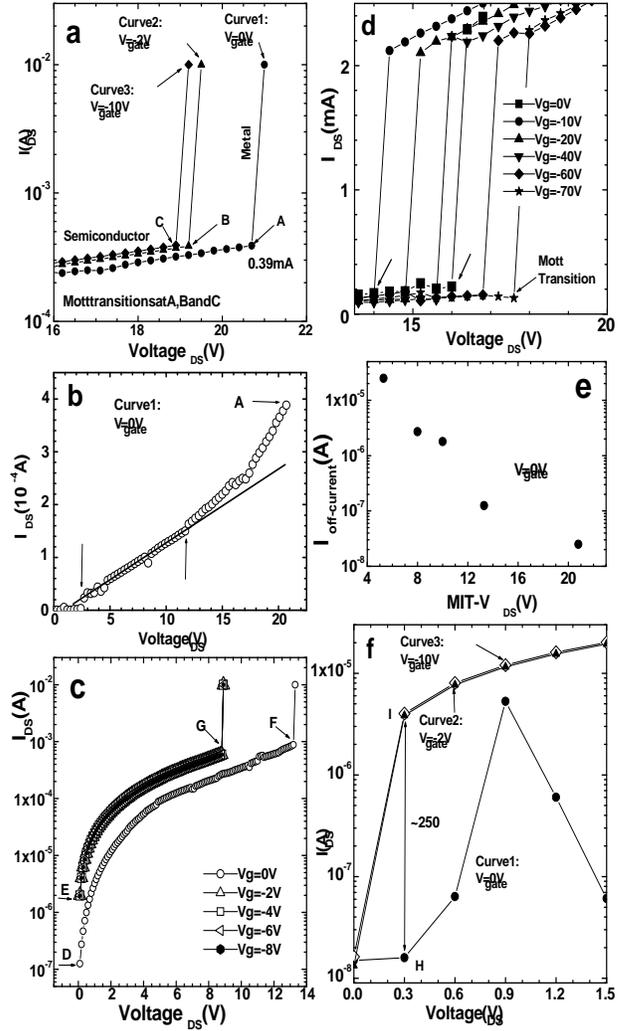}}
\vspace{0.5cm} \caption{${\bf a}$, $I_{DS}$ vs $V_{DS}$ of
transistor 1. The abrupt MIT occurs at A, B, and C. At the jumped
point in curve 1, the current density is about $J$=3$\times$10$^5$
A/cm$^2$, which is current collective motion observed in metal.
${\bf b}$, $I_{DS}$ vs $V_{DS}$ below the MIT point A of curve 1.
The Ohmic behavior is shown from $V_{DS}$=2.5V up to 12V. ${\bf
c}$, $I_{DS}$ vs $V_{DS}$ of transistor 2. ${\bf d}$, $I_{DS}$ vs
$V_{DS}$ near the abrupt MIT of transistor 3.  Above $I_{DS}$=2mA,
the Ohmic behavior is exhibited. At the jumped point, current
density is $j$=0.9$\times$10$^5$ A/cm$^2$. ${\bf e}$, Off-current
vs MIT-$V_{DS}$. The off-currents are extracted from 5
transistors. ${\bf f}$, $I_{DS}$ vs $V_{DS}$, magnified below
$V_{DS}$=1.5V in Fig. 3a. Black diamonds are data measured at
$V_{gate}$=-2V.}
\end{figure}

\end{document}